\documentclass[aps,prd,reprint,showpacs,nofootinbib,superscript address]{revtex4-2}
\usepackage[utf8]{inputenc}
\usepackage{scalerel,stackengine}
\usepackage{physics}
\usepackage{tocbasic}
\usepackage{amssymb}
\usepackage{upgreek}
\usepackage{hhline}
\usepackage{amsmath}
\usepackage{mathtools}
\usepackage[dvipsnames]{xcolor}
\usepackage{multirow,tabularx}
\usepackage{siunitx}
\usepackage{multirow}
\usepackage{graphicx}
\usepackage{etoolbox}
\usepackage{notoccite}
\usepackage{natbib}
\usepackage{mathrsfs}
\usepackage{tensor}
\usepackage{accents}

\usepackage{tikz}
\usetikzlibrary{arrows.meta, positioning}

\DeclareTOCStyleEntry[numwidth=20pt,linefill=\bfseries\TOCLineLeaderFill]{tocline}{section}
\DeclareTOCStyleEntry[entryformat=\textit,numwidth=10pt,linefill=\TOCLineLeaderFill]{tocline}{subsection}
\DeclareTOCStyleEntry[entryformat=\textit,numwidth=10pt,linefill=\TOCLineLeaderFill]{tocline}{subsubsection}

\usepackage{hyperref}
\hypersetup{%
     colorlinks = true,%
     linkcolor = Blue,%
     citecolor = Blue,%
     filecolor = Blue,%
     urlcolor = Blue%
     }%
\usepackage[capitalize,nameinlink]{cleveref}
\crefname{paragraph}{paragraph}{paragraphs}
\Crefname{paragraph}{Paragraph}{Paragraphs}

\begin{document}

\title{Gravitational mass generation and consistent non-minimal couplings:\\ cubics and quartics of a massive vector}

\author{Carlo Marzo}
\email{carlo.marzo@kbfi.ee}
\affiliation{Laboratory for High Energy and Computational Physics, NICPB, R\"{a}vala 10, Tallinn 10143, Estonia}

\begin{abstract}
An attempt to evade the strict uniqueness of consistent interactions involving spin-2 particles is made by modifying the Noether procedure from the outset. A vector field is introduced, coupled to a graviton already at the level of quadratic mixing. The byproduct is a gauge-invariant mass for the vector and novel consistent interactions, here derived and tested up to quartic order. A simple geometric interpretation of the vector field appears possible.
\end{abstract}

\maketitle


\section{Introduction}
The quantum-field approach to explore the dynamic of spin-2 particles is not just a redundant way to re-obtain the known large scale properties of the gravitational interaction. It also offers an opportunity to reconsider some of the established geometrical features as emergent, and to develop the theory bottom-up, with a closer focus to the typical quantum consistency requirements of unitarity and (generic) renormalizability. The latter, a fundamental prerequisite for a theory to have any predictive use \cite{Gomis:1995jp,Donoghue:1993eb,Donoghue:1994dn,Gasser:1983yg,Entem:2003ft}, is too easily given away under geometrically-justified deformations of the minimal Einstein-Hilbert setup \cite{Marzo:2021iok,Marzo:2024pyn,Barker:2025xzd,Barker:2025rzd,Barker:2025fgo}. Conversely, starting from a healthy free-theory of spin-2 particles and building up consistent interactions, while possibly less elegant, builds up to, by construction, either predictive computational frameworks or nothing at all\cite{Fang:1978rc,Deser:1963zzc,Berends:1984rq,Boulanger:2000rq,Barnich:2017nty,Zinovev:2017cdp}. The second outcome manifesting whenever, for a given field content, no consistent interactions can be obtained \footnote{Trivial interactions obtained by squaring an the invariant kinetic term can always be presented. Accounting for such possibility, we proceed by only focusing on non-trivial interactions, and our claim on 'non-existence' must be understood under such stricter definition.}. By adopting this more \emph{fundamental} procedure, we can focus on possible new avenues which might be invisible once the geometrical embedding is accepted. This program is not yet complete, nor successful, and we do not claim to have anything more than a few promising results along with some new modeling opportunities that might have eluded the attention of the community.  
Moreover, this is not an original attempt either. The closest to our approach can be found in the work of Wald and Cutler\cite{Wald:1986bj,Cutler:1986dv}, on Yang-Mills-inspired self-interacting spin-2. The case of Wald is particularly interesting given that, reliant on consistent interactions, considered an internal index to deform the theory at the cubic level onward. The impossibility of doing so by preserving the unitary structure of the kinetic term \cite{Boulanger:2000rq}, as well as the lack of a consistent continuation to quartic vertices, points to a more drastic attempt: \emph{field mixing}. 
That extra fields might play a facilitating role \emph{already at the quadratic level} is plausible and, possibly, unavoidable. For instance, it is not even possible to consider a cubic renormalizable ansätz for self-interactions (in the symbolic format described in the Appendix: $\mathsf{H}^3$) as long as the gauge symmetry is the derivative of a vector $\mathsf{\xi}$: $\delta^0 \mathsf{H} = \partial \mathsf{\xi}$. This can be easily seen by the fact that gauge invariance would require cancellation of the class $\mathsf{H}^2 \partial \mathsf{\xi}$ from the gauge deformation of the kinetic term, a 2-derivative object to start with. We will begin our investigation considering a class of theories with a rank-2 symmetric field $\mathsf{H} \equiv H_{ab} = H_{(ab)}$ mixed, quadratically, with a rank-1 field $\mathsf{V} \equiv V_{a}$.  
We do so by finding quadratic models with a healthy spectrum completely shaped by mild deformations of linear diffeomorphism invariance. In other words, no tuning is imposed to remove ghosts, and the model is ready for a consistent analysis of its possible completions.   
The key to triggering a deviation from the interplay between vectors and gravitons, as strictly governed by nonlinear diffeomorphism, lies in employing the trace of $\mathsf{H}$ as a portal. Consequently, the vector field must respond differently under the longitudinal component of diffeomorphism and, in order to enter the theory already at the quadratic level, must transform with an inhomogeneous shift. 
It is perhaps unsurprising that the trace of $\mathsf{H}$ can facilitate non-trivial interactions while retaining an auxiliary role. A precedent for this already appears in the original Fierz–Pauli construction, where the field equation for a traceless rank-2 field was found to require an auxiliary lower-rank field in order to admit a consistent Lagrangian formulation and support interactions \cite{Fierz:1939ix}. 
A bottom-up approach affords greater freedom to explore different ways of mixing the two fields under a common gauge symmetry. A recurring outcome of this exploration is the emergence of massive states within a gauge-invariant Lagrangian, a familiar phenomenon, already realized in the Stückelberg construction for lower-rank fields. Our analysis reveals, broadly, two distinct ways of constructing intertwined transformation rules for the two fields. The more radical of these yields a gauge-invariant description of massive spin-2 states, and is the subject of a forthcoming work. Here, instead, we study the minimal modification, which generates a mass for the vector field while preserving massless spin-2 propagation. Interestingly, regardless the geometric-agnostic approach, the resulting gauge transformation for the vector will closely resemble the analogous one of a (contraction of the) affine connection in curved space.

\section{The Noether procedure}
The mismatch between the physical concept of particles and the local fields introduced for their description naturally leads to the introduction of gauge symmetries. In general, the multiple particle sectors carried by a given field impose mutually incompatible requirements on the parameter space for the theory to achieve unitary and causal propagation. The symmetric rank-2 tensor $\mathsf{H}$ provides a paradigmatic example, as it carries the irreducible Wigner components
\begin{align}\label{eq1}
\mathsf{H} \supset\; 2_H^+ \oplus 1_H^- \oplus 0_{H_1}^+ \oplus 0_{H_2}^+\,.
\end{align}
Following a broadly used notation \cite{Percacci:2020ddy}, we labeled each sector with its spin $S$ and parity $P$ eigenvalues as $S^P$, adding an additional subscript to identify the parent field and distinguish between sectors of the same spin. Similarly, the vector representation $\mathsf{V}$ carries the more minimal set of components 
\begin{align}\label{eq2}
\mathsf{V} \supset\; 1_V^- \oplus 0_{V}^+\,.
\end{align}
The generic quadratic Lagrangian $\mathcal{L}^{(0)}$ for $\mathsf{H}$ alone is encoded in the following monomials 
\begin{align} \label{monomials}
\mathsf{\partial^2 H^2} \equiv \; 
& k_1 \, H^{ab} \partial_a \partial_b H^{c}{}_{c}
+ k_2 \, H^{ab} \partial_c \partial_b H_{a}{}^{c}  \notag \\
& + k_3 \, H^{a}{}_{a} \Box H^{b}{}_{b} + k_4 \, H^{ab} \Box H_{ab}
 \, , \notag \\  \notag \\ 
\mathsf{H^2} \equiv \; 
& hh_1 \, H^{ab} H_{ab} 
+ hh_2 \, H^{a}{}_{a} H^{b}{}_{b}\, .
\end{align}
A known spectral analysis \cite{vanderBij:1981ym} (simple to recover independently via publicly available tools \cite{Barker:2024juc}) illustrates the incompatibility between the $2_H^+$ and $1_H^-$ sectors. The issue is resolved by demanding invariance under the following local shift
\begin{align} \label{tdiffeo}
    \delta^0 H_{ab} = \partial_{(a} \xi^T_{b)} \, ,\,\, (\partial_{a} \xi^{T a} = 0) 
\end{align}
which consistently removes the three-states of the $1^-$ sector. The quadratic Lagrangian invariant under the transversal shift is non-pathological but overabundant, due to the extra presence of a massive scalar state, on top of the massless spin-2. A natural step is then to promote the gauge generator to a full vector 
\begin{align} \label{diffeo}
 \delta^0 H_{ab} = \partial_{(a} \xi_{b)} \, ,\,\,  
\end{align}
which shapes uniquely (within nuances detailed in the next section) the Einstein-Hilbert Lagrangian. 

The identification of a gauge symmetry for the generic quadratic Lagrangian $\mathcal{L}^{(0)}$ goes hand in hand with a degeneracy \emph{identity} of the corresponding kinetic term 
\begin{align} \label{eqNoe0}
\delta^{(0)}_{\xi}\, \mathcal{L}^{(0)} = 0 \, .
\end{align}
Notably, Eq.(\ref{eqNoe0}) holds for any field configuration and not just for the solutions of the corresponding linear problem. This identity is the basis of a powerful self-consistent procedure to bootstrap the full interacting theory as a perturbative expansion in powers of (weak) fields. Introducing a weak coupling $g$ to track the expansion order, we can consider the full Lagrangian as a series of increasing polynomial degree
\begin{align}
\mathcal{L} = \mathcal{L}^{(0)} + g\, \mathcal{L}^{(1)} + g^2\, \mathcal{L}^{(2)} + g^3\, \mathcal{L}^{(3)}  \cdots
\end{align}
with a parallel structure for the gauge transformation
\begin{align}
\delta_{\xi} = \delta^{(0)}_{\xi} + g\, \delta^{(1)}_{\xi} + g^2\, \delta^{(2)}_{\xi} + g^3\, \delta^{(3)}_{\xi} + \cdots
\end{align}
Demanding invariance of the full Lagrangian under the full gauge transformation becomes non-trivial and intertwined due to Eq.(\ref{eqNoe0}). For the scope of this paper, we only need the identities shaping the $\mathcal{O}(g)$ 
\begin{align}
\delta^{(0)}_{\xi}\, \mathcal{L}^{(1)} + \delta^{(1)}_{\xi}\, \mathcal{L}^{(0)} = 
\partial_{\mu} \mathcal{J}^{\mu}_{(1)}
\end{align}
and $\mathcal{O}(g^2)$ order
\begin{align}
\delta^{(0)}_{\xi}\, \mathcal{L}^{(2)} + \delta^{(1)}_{\xi}\, \mathcal{L}^{(1)} 
+ \delta^{(2)}_{\xi}\, \mathcal{L}^{(0)} = \partial_{\mu} \mathcal{J}^{\mu}_{(2)}
\end{align}
which defines the consistent cubic and quartic vertices of the interacting Lagrangian. 

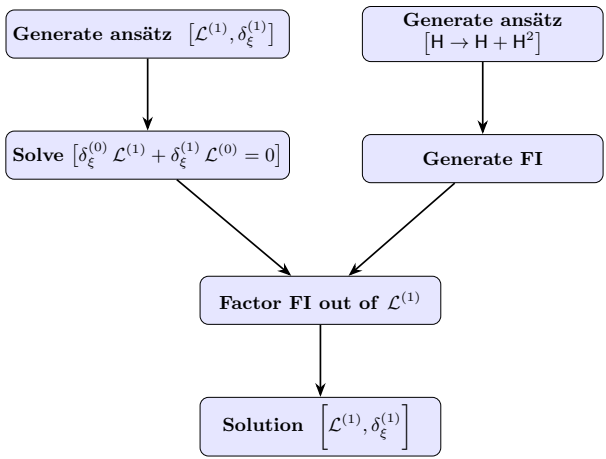
\begin{figure}
\scalebox{0.8}{
\begin{tikzpicture}[
    node distance = 1.2cm,
    box/.style = {
        rectangle,
        rounded corners = 4pt,
        draw = black,
        fill = blue!10,
        minimum width = 4cm,
        minimum height = 0.8cm,
        align = center,
        font = \small
    },
    arrow/.style = {
        -Stealth,
        thick
    }
]
\node[box] (step1) {\textbf{Generate ansätz } $\big[\mathcal{L}^{(1)},\delta^{(1)}_{\xi}\big]$ };
\node[box, below = of step1] (step2) {\textbf{Solve} $\big[ \delta^{(0)}_{\xi}\, \mathcal{L}^{(1)} + \delta^{(1)}_{\xi}\, \mathcal{L}^{(0)} = 0 \big]$};
\node[box, below right=1.6cm and -1.5cm of step2] (step3) {\textbf{Factor FI out of }$\mathcal{L}^{(1)}$};

\node[box, right = of step1] (step2b) {\textbf{Generate ansätz} \\ $\big[ \mathsf{H} \rightarrow \mathsf{H} + \mathsf{H}^2 \big]$ };
\node[box, below = of step2b] (step3b) {\textbf{Generate FI}};
\node[box, below = of step3] (step4b) {\textbf{Solution } $\bigg[\mathcal{L}^{(1)},\delta^{(1)}_{\xi}\bigg]$ };

\draw[arrow] (step1) -- (step2);
\draw[arrow] (step2) -- (step3);
\draw[arrow] (step3) -- (step4b);

\draw[arrow] (step3b) -- (step3);

\draw[arrow] (step2b) -- (step3b);

\end{tikzpicture}
}
    \caption{The algorithm to isolate cubic candidate interactions and gauge deformations, factoring fake-interactions (FI) out. The next step for quartic vertices follows a similar logic.}
    \label{fig:NoAlgo}
\end{figure}

This \emph{Noether} procedure lends itself naturally to an algorithmic implementation. Refined versions can be found in the literature, largely based on an equivalent cohomological reformulation of the symmetries \cite{Henneaux:1989jq,Barnich:2025jna,Barnich:2000zw,Henneaux:1997bm}. Our approach follows closely the Noether machinery presented above. The algorithmic implementation of the process requires only a functional derivative with a kernel in the space of boundary terms, together with a tensor canonicalizer, both provided by the \texttt{xAct} package\footnote{\texttt{http://www.xact.es/}}\cite{Nutma:2013zea,Brizuela:2008ra}. Beyond these technical tools, additional (back)checks must be introduced to guard against false positives. The most important is a careful accounting of \emph{trivial} couplings, or fake interactions (FI), that can be removed through a field redefinition acting on the kinetic term. Fig.(\ref{fig:NoAlgo}) illustrates the steps we found necessary to isolate the cubic vertex candidate. The subsequent quartic order follows an analogous chain.
\section{Bootstrapping Gravity} \label{BootGR}
Before tackling the vector-tensor mixed system, we validate the implemented routines against the familiar case in which $\mathsf{H}$ transforms as in Eq.~(\ref{diffeo}) while $\mathsf{V}$ shifts as
\begin{align} \label{qed}
 \delta^0 V_{a} = \partial_{a} \omega\, ,\,\,
\end{align}
as required to consistently decouple the ghost-like $0^+_V$ component. This scenario is \emph{unmixed}, in the sense that the two fields carry independent gauge symmetries. Each field will nonetheless, be allowed to respond \emph{homogeneously} to the gauge transformation of the other. 
We initiate the Noether procedure providing a generic and unbiased parameterization of the cubic Lagrangian $\mathcal{L}^{(1)}$
including all terms up to two-derivatives:
\begin{align} \label{L1Diffeo}
 \mathcal{L}^{(1)} &= \mathsf{H}^3 + \partial \mathsf{V}^3 +  \mathsf{H} \mathsf{V}^2 + \mathsf{H}^2 \partial \mathsf{V} +  \partial^2 \mathsf{H}^3 +  \mathsf{H} \partial^2 \mathsf{V}^2 \, . \notag \\
\end{align}
To fully solve for the $\mathcal{O}(g)$ gauge constraint, we need the ansätz for the gauge deformations as a function of the local parameter $\xi_{a}$ and $\omega$. Once a decision is made for the $\mathcal{L}^{(1)}$ ansätz, the operators entering the gauge deformations are completely determined (\emph{vice versa}, we can select the gauge deformation first and derive the form of the interaction operators later): 
\begin{align} \label{GaugeDeform}
 \delta^1_{\xi} \mathsf{H}  &= \mathsf{H} \partial \xi +  \mathsf{V} \xi \,, \quad   \delta^1_{\xi} \mathsf{V}  = \mathsf{V} \partial \xi +  \mathsf{H} \xi \, , \nonumber \\
 \delta^1_{\omega} \mathsf{H}  &= \mathsf{H} \omega + \mathsf{H} \partial^2 \omega +  \mathsf{V} \partial \omega \, , \nonumber \\
 \delta^1_{\omega} \mathsf{V}  &= \mathsf{V} \omega + \mathsf{V} \partial^2 \omega +  \mathsf{H} \partial \omega \, . 
\end{align}
\emph{Do these assumptions suffice to uniquely recover the full non-linear structure of GR? }\\\\ 
The question of whether GR can be reconstructed from quantum consistency alone has been debated at length, with constructive examples on one side and objections over the biased selection of candidate interaction terms on the other \cite{Kraichnan:1955zz,Gupta:1954zz,Padmanabhan:2004xk,Deser:1969wk,Butcher:2009ta}. We bring no final evidence for what concerns the full non-perturbative reconstruction and confine our comments to the unambiguous product of our $\mathcal{O}(g^2)$ computation.
The consistency requirements prove strong enough that, in Eq.(\ref{L1Diffeo}), only specific combinations of $\partial^2 \mathsf{H}^3$ and $\mathsf{H} \partial^2 \mathsf{V}^2$ survive, with coefficients that, within each set, precisely reproduce those obtained by general covariantization\footnote{In this procedure, the field $\mathsf{H}$ is identified with the metric fluctuation around Minkowski space, $g_{ab} = \delta_{ab} + \kappa H_{ab}$.}. However, the cubic analysis alone seems insufficient to capture the universal and geometric character of the interaction: it permits independent overall coefficients multiplying $\partial^2 \mathsf{H}^3$ and $\mathsf{H} \partial^2 \mathsf{V}^2$ separately. Universality can in principle be restored by supplementing the analysis with an external consistency check, such as Weinberg's low-energy theorems \cite{Weinberg:1964ew,Weinberg:1965rz}. However, our aim is for the algorithm to be self-contained, and we therefore proceed to the $\mathcal{O}(g^2)$ order without further changes. The ansätze for the $\mathcal{O}(g^2)$ analysis then follow straightforwardly
\begin{align} \label{L2Diffeo}
 \mathcal{L}^{(2)} &= \mathsf{H}^4 + \mathsf{V}^4+ \partial \mathsf{H}  \mathsf{V}^3 + \mathsf{H}^2 \mathsf{V}^2 \notag \\
 & + \mathsf{H}^3 \partial \mathsf{V} +  \partial^2 \mathsf{H}^4 +  \mathsf{H}^2 \partial^2 \mathsf{V}^2 \, , \notag \\
\end{align}
as do the generic gauge transformations
\begin{align} \label{GaugeDeform4}
 \delta^2_{\xi} \mathsf{H}  &= \mathsf{H}^2 \partial \xi +  \mathsf{V} \mathsf{H} \xi +  \partial \mathsf{V}^2 \mathsf{H} \xi \,,
 \nonumber \\
 \delta^2_{\xi} \mathsf{V}  &= \mathsf{V} \mathsf{H} \partial \xi + \mathsf{H}^2 \xi + \mathsf{V}^2 \xi \, , \nonumber \\
 \delta^2_{\omega} \mathsf{H}  &= \mathsf{H}^2 \omega + \mathsf{H}^2 \partial^2 \omega +  \mathsf{V} \mathsf{H} \partial \omega + \mathsf{V}^2 \omega + \mathsf{V}^2 \partial^2 \omega \, , \nonumber \\
 \delta^2_{\omega} \mathsf{V}  &= \mathsf{V}^2 \partial \omega + \mathsf{H}^2 \partial \omega +  \mathsf{V} \mathsf{H} \omega +  \mathsf{V} \mathsf{H} \partial^2 \omega \, . 
\end{align}
Obviously, most of them appear misplaced for the problem in question, but a general and unbiased approach must be maintained for less conservative studies we intend to carry. Here, we find that a non-trivial solution exists at this order only if the two independent parameters accompanying $\partial^2 \mathsf{H}^3$ and $\mathsf{H} \partial^2 \mathsf{V}^2$ are forced to coincide, thereby recovering universality. 
\section{Tensor-Vector beyond GR} 
The previous test (re-)confirm the rigidity of linear diffeomorphism: once Eq.~(\ref{diffeo}) and Eq.~(\ref{qed}) are taken as the seed of the Noether procedure, no deviation from the known gravitational interactions can emerge. The bottom-up approach, however, points to an alternative path. As anticipated, new possibilities are unlocked by considering gauge transformations that act on $\mathsf{H}$ and $\mathsf{V}$ simultaneously.
While many such transformations can in principle be entertained, only a restricted class admits a ghost-free quadratic model. Within this class, we identify two qualitatively distinct routes distinguished by how the vector field responds to the transversal shift Eq.(\ref{tdiffeo}). 
One natural possibility for mixed scenarios is to build on Eq.~(\ref{tdiffeo}) by assigning the vector the parallel shift $\delta^0 V_{\mu} = \xi^T_{\mu} $. Under this choice, the vector relinquishes the propagation of its $1^-$ sector, and the theory flows towards a description of massive spin-2 states. The bottom-up investigation of this branch is reserved for a forthcoming companion paper.

Here, we instead focus on the consequences of allowing the vector field to retain its propagating role. This requirement rules out the transverse diffeomorphism as a candidate shared symmetry, and the remaining natural candidate can only be generated by a scalar parameter. Among the various possibilities, a particularly transparent and embarrassingly simple one emerges which employs the trace of $\mathsf{H}$ as the scalar component of a Stückelberg realization:
\begin{align} \label{Procadiffeo}
 \delta^0 H_{ab} = \partial_{(a} \xi_{b)} \, ,\,\,\,\,\,\,  
 \delta^0 V_{a} = M^{-1} \partial_{a} (\partial \cdot \xi) \, ,\,\, 
\end{align}
with $M$ a dimensionful free parameter ($[M] = \text{mass}$).
This symmetry must be imposed onto the generic quadratic Lagrangian of the mixed tensor-vector system,
obtainable extending the set in Eq.(\ref{monomials}) with 
\begin{align} \label{monomialsVH}
\mathsf{\partial^2 V^2} \equiv \; 
& v_1 \, V^{a} \partial_a \partial_b V^{b}
+ v_2 \, V^{a} \Box V_{a}  , \notag \\  
\mathsf{V^2} \equiv \; 
& vv_1 \, V^{a} V_{a}\, ,   \notag \\
\mathsf{H} \partial \mathsf{V} \equiv \; 
& hv_1 \, H_{a}{}^{a} \partial_{b} V^{b} + \frac{hv_2}{2} \, H_{ab} \big(\partial_{a} V^{b} + \partial_{b} V^{a}\big)\, . \notag \\
\end{align}
Then, demanding invariance, we obtain
\begin{align} \label{lagVH}
\mathcal{L}_{VH} = \; 
& + v_1 \, \big( V^a \partial_b \partial_a V^b
- \, V^a \Box V_a \big) \notag \\
& + hv_1 \big(\, M^2 \, V_a V^a 
+ M\, H^{a}{}_{a} \, \partial_b V^b - \frac{1}{4} \, H^{a}{}_{a} \, \Box H^{b}{}_{b}\big) \notag \\
& + k_1 \, \big( H^{ab} \partial_a \partial_b H^{c}{}_{c}
- \, H^{ab} \partial_c \partial_b H_{a}{}^{c} \notag \\
& + \frac{1}{2} \, H^{ab} \Box H_{ab} - \frac{1}{2} \, H^{a}{}_{a} \, \Box H^{b}{}_{b} \big) \, , 
\end{align}
where we rescaled the dimensionful parameter by opportune powers of $M$. The result is very simple and almost trivial given that it is precisely made up of a massless Fierz-Pauli combination, a Maxwell term and a Stückelberg completion with the trace of $\mathsf{H}$; each combination modulated, respectively, by the free parameters $k_1$, $hv_1$ and $v_1$. A direct spectral analysis \cite{Barker:2024juc} confirms that the model admits, with no strict tuning, a unitary propagation of a graviton and massive spin-1 state as long as $v_1 < 0\land hv_1 > 0 \land k_1 < 0$. To conclude this section: we proved that a rank-2 tensor propagating a massless spin-2 particle can, with no (radiatively unstable) tuning, generate a gauge-invariant mass term for a vector particle.

\section{Bootstrapping New Interactions}
\subsection{Empty solutions and homogeneous gauge transformations}
For the identification of a quadratic, unitary, and gauge-invariant model to be of any use, it must serve as the foundation of a non-linear theory. Our goal is therefore to apply to the Lagrangian Eq.~(\ref{lagVH}) the same Noether algorithm tested in the familiar context of GR.
Before doing so, we pause to stress an important point: the possible absence of non-trivial interactions emerging from the Noether procedure in mixed systems does not constitute a no-go theorem for their existence in general. It may instead indicate that the natural completion of the theory requires the introduction of external matter, transforming homogeneously under one or more of the gauge symmetries. The paradigmatic example is, once again, the classical Stückelberg model.
\begin{align} \label{lagSV}
\mathcal{L}_{SV} = \; 
& + v_1 \, \big( V^a \partial_b \partial_a V^b
- \, V^a \Box V_a \big) \notag \\
& + hv_1 \big(\, M^2 \, V_a V^a 
- 2 M\, \partial_b S \,  V^b - \, S \, \Box S \big)  
\end{align}
obtained by imposing 
\begin{align} \label{TradStuck}
 \delta^0 S = M \omega \, ,\,\,\,\,\,\,  
 \delta^0 V_{a} = \partial_{a} \omega \, .\,\, 
\end{align}
Had we seeded the Noether procedure with Eq.(\ref{TradStuck}) and $\mathcal{L}^{(0)} \equiv \mathcal{L}_{SV}$, the search for interactions would have yielded no solution. However, abandoning the mixed-symmetry setup of Eq.(\ref{lagSV}) at that stage would have missed a complete theory. In such a theory, the field $S$ is accompanied by a second real scalar $h$ which transforms \emph{homogeneously} and therefore does not mix, together forming a complex representation $\phi$ of the abelian $U(1)$ group $\phi = v(M) + h + i S $. This is  the familiar non-linear realization at the core of the spontaneous symmetry breaking with a vacuum $v(M)$. Correspondingly, the inhomogeneously shifting field $S$ is identified as the Goldstone boson.
\subsection{Cubics to Quartics}
Notwithstanding that additional fields might facilitate the search for a non-linear completion, we nevertheless investigate whether already the minimal content of Eq.~(\ref{lagVH}), endowed with the gauge symmetry Eq.~(\ref{Procadiffeo}), could sustain a consistent deformation up to $\mathcal{O}(g^2)$. 
This search adopts the same ansätze used to uniquely recover the standard gravitational interactions of Section~\ref{BootGR} (with the simplification of not having an independent $\omega$ parameter). As in that case, candidates survive the consistency checks of the Noether procedure at the cubic level, even after accounting for the removal of the FI, an outcome that, while not guaranteed, is not unexpected. Somewhat surprisingly, one combination of cubics survives the tighter constraints generated by the $\mathcal{O}(g^2)$ invariance with a non-trivial deformation of the gauge symmetry. It is this particular (and unique, if we require interactions to start at cubic level) solution which we present here.

Once we have demonstrated the existence of unitary regions for the parameter space, we can normalize the terms of the quadratic Lagrangian Eq.~(\ref{lagVH})
\begin{align}
\mathcal{L}^{(0)} = \;
&\frac{M^2}{2}\,V_a V^a
+ \frac{M}{2}\,H^b{}_b\,\partial_a V^a \notag \\
&- \frac{1}{2}\,H^{ab}\partial_b\partial_a H^c{}_c
+ \frac{1}{2}\,H^{ab}\partial_c\partial_b H_a{}^c
- \frac{1}{4}\,H^{ab}\Box H_{ab}\notag \\
&
+ \frac{1}{8}\,H^a{}_a\,\Box H^b{}_b - \frac{1}{2}\,V^a\partial_b\partial_a V^b
+ \frac{1}{2}\,V^a\Box V_a
\end{align}
and proceed with the Noether algorithm.
It is apparently surprising that, regardless the starkly different beginnings offered by Eq.~(\ref{lagVH}) and the general ansätz Eq.~(\ref{GaugeDeform}), the surviving consistent deformation results to be 
\begin{align} \label{surprise}
\delta^{(1)} H_{ms} = \;
g \Bigl[
&\xi^a \partial_a H_{ms}
+ H_{as}\,\partial_m \xi^a
+ H_{am}\,\partial_s \xi^a
\Bigr] \, , \notag \\
\delta^{(1)} V_m = \;
g \Bigl[
&\xi^a \partial_a V_m
+ V^a \partial_m \xi_a
\Bigr] \, , 
\end{align}
a Lie derivative under the shift $\xi$. Admittedly, Eq.~(\ref{surprise}) is not the raw output of the Noether procedure. As discussed, the solution is defined up to a number of redundancies and field redefinitions, which can be exploited to simplify its form. In particular, the fact that the vector transformation immediately took the form of a Lie derivative suggested that such reduction was achievable also for the rank-2 tensor, whose original output was slightly more involved. 

The coupling $g$ in Eq.~(\ref{surprise}) carries dimension $[{\rm \text{mass}}]^{-1}$, and a rescaling by the vector mass scale $M$ might appear natural. However, we see no reason to parametrically tie two independent scales together, and leave $g$ as it is. 

The cubic interaction\footnote{Formatting of long equations is prone to typos. We refer the reader to \texttt{https://github.com/CarloO3/MaterialVecGrav} as the container for raw, digital versions of the forthcoming formulas.} consistent with this first-order deformation also contains some unusual features, tied to the novel presence of mass and mixing terms in Eq.~(\ref{lagVH}). Quite generally, even restricting a comparison only to terms $\mathsf{\partial^2 H^3}$, the cubics found \emph{cannot be matched to those of GR}:
{\small
\begin{widetext}
\begin{align}
&\mathcal{L}^{(3)} = \;
g \Bigl\{
M^2 \Bigl[
\frac{1}{4}\,H^n{}_n V_m V^m
- \frac{1}{2}\,H_{mn} V^m V^n
\Bigr] \notag \\[8pt]
&+ M \Bigl[
- \frac{1}{4}\,H^n{}_n V^m\,\partial_a H_m{}^a
+ \frac{1}{4}\,H^m{}_n V_m\,\partial^n H^a{}_a
- \frac{1}{4}\,H^a{}_a H_{mn}\,\partial^n V^m 
- \frac{1}{4}\,H_{na} H^{na}\,\partial_m V^m
+ \frac{1}{8}\,H^a{}_a H^n{}_n\,\partial_m V^m
\Bigr] \notag \\[8pt]
&+ \frac{1}{2}\,H_{na}\,\partial^a V_m\,\partial^n V^m
- H_{ma}\,\partial^a V_n\,\partial^n V^m
+ \frac{1}{4}\,H^a{}_a\,\partial_m V_n\,\partial^n V^m
+ \frac{1}{2}\,H_{ma}\,\partial_n V^a\,\partial^n V^m
- \frac{1}{4}\,H^a{}_a\,\partial_n V_m\,\partial^n V^m \notag \\[8pt]
&+ \frac{1}{3}\,H_{ma} H^{mn}\,\partial^a\partial_n H^b{}_b
- \frac{1}{6}\,H^m{}_m H^{na}\,\partial_a\partial_n H^b{}_b
+ \frac{1}{6}\,H_{mn} H^{mn}\,\partial_b\partial_a H^{ab}
- \frac{1}{12}\,H^m{}_m H^n{}_n\,\partial_b\partial_a H^{ab} \notag \\
&+ \frac{1}{3}\,H^{ab} H^{mn}\,\partial_b\partial_a H_{mn}
- \frac{2}{3}\,H_{ma} H^{mn}\,\partial_b\partial^a H_n{}^b
+ \frac{1}{2}\,H^m{}_m H^{na}\,\partial_b\partial_a H_n{}^b
- \frac{1}{12}\,H_{mn} H^{mn}\,\partial_b\partial^b H^a{}_a \notag \\
&+ \frac{1}{24}\,H^m{}_m H^n{}_n\,\partial_b\partial^b H^a{}_a
+ \frac{1}{3}\,H_{ma} H^{mn}\,\partial_b\partial^b H_n{}^a
- \frac{1}{6}\,H^m{}_m H^{na}\,\partial_b\partial^b H_{na}
- \frac{1}{3}\,H^{ab} H^{mn}\,\partial_b\partial_n H_{ma} \notag \\[8pt]
&- \frac{1}{12}\,H^{mn}\,\partial_a H^b{}_b\,\partial^a H_{mn}
+ \frac{1}{48}\,H^m{}_m\,\partial_a H^b{}_b\,\partial^a H^n{}_n
+ \frac{1}{6}\,H^{mn}\,\partial^a H_{mn}\,\partial_b H_a{}^b
- \frac{1}{12}\,H^m{}_m\,\partial^a H^n{}_n\,\partial_b H_a{}^b \notag \\
&+ \frac{1}{12}\,H^{mn}\,\partial_m H^{ab}\,\partial_n H_{ab}
- \frac{1}{24}\,H^{mn}\,\partial_m H^a{}_a\,\partial_n H^b{}_b
+ \frac{1}{6}\,H^{mn}\,\partial_a H^b{}_b\,\partial_n H_m{}^a
- \frac{1}{3}\,H^{mn}\,\partial_b H^{ab}\,\partial_n H_{ma} \notag \\
&+ \frac{1}{6}\,H^m{}_m\,\partial_b H^{ab}\,\partial_n H^n{}_a
- \frac{1}{6}\,H^{mn}\,\partial_a H_{nb}\,\partial^b H_m{}^a
+ \frac{1}{6}\,H^{mn}\,\partial^b H_{na}\,\partial_b H_{mb}
+ \frac{1}{12}\,H^m{}_m\,\partial_a H_{nb}\,\partial^b H^{na}
- \frac{1}{24}\,H^m{}_m\,\partial^b H_{na}\,\partial_b H^{na}
\Bigr\}
\end{align}
\end{widetext}
}
The emergence of dimension-3 and dimension-4 operators in the cubic Lagrangian is an expected feature of the mixed system with quadratic terms. 
This is a good point to pause and present the ansätze we used for the field redefinitions, given it might point to some neglected terms on our side:
\begin{align}
&    \mathsf{H} \to \mathsf{H} + \mathsf{H}^2 + \mathsf{V}^2 + \partial \mathsf{V}\mathsf{H} \, , \notag \\
&    \mathsf{V} \to \mathsf{V} + \mathsf{V} \mathsf{H} + \partial \mathsf{H}^2 + \partial \mathsf{V}^2 \, .
\end{align}
Notice that, for precaution and considering possible cancellations, we have allowed 1-derivative terms even though they will, in general, contribute to 3-derivative cubic FI. 
Overall, we stress that the final \emph{essential} form of the cubic interaction suggests neglected opportunities for searching properly renormalizable setups within mixed systems involving more clever structures than the one we used. The interconnection between operators of different dimensionality can be further elucidated by an analysis of the induced quantum corrections, a study we are currently pursuing.

It is a common fate for many promising cubic interactions and corresponding gauge deformations to encounter an obstruction when continuing to quartic order \cite{Barnich:1993vg,Deser:1990bk}. As mentioned, this is not the case for this solution as we have explicitly verified at $\mathcal{O}(g^2)$ order.
Pleasantly, the Lie derivative structure for the vector in Eq.(\ref{surprise}) is not modified by any high-order deformation $ \delta^{(2)} V_{m} = 0 $. A slightly more involved \emph{raw} result is, again, produced for the deformation of the rank-2 tensor
\begin{align}
&\delta^{(2)} H_{ms} = \;
 \bigl(g^2 +  C_1\bigr) H^b{}_b H_{ms}\,\partial_a \xi^a \notag \\
&+ \bigl(g^2 + C_2\bigr) H_{ab} H_s{}^b\,\partial^a \xi_m  + \bigl(g^2 + C_3\bigr) H_{ab} H_m{}^b\,\partial^a \xi_s \notag \\
&+ \cdots\,
\end{align}
Each $C_i$ is a combination of free parameters drawn from the definitions introduced in Eqs.~(\ref{L2Diffeo}--\ref{GaugeDeform4}), and each represents an independent and consistent, albeit likely trivial, solution of the Noether procedure. Since we are interested only in deformations directly connected to the cubic vertex found above, we could in principle restrict our attention to the part proportional to $g^2$, setting the $C_i$ consistently to zero. However, motivated by the geometric interpretation of $\mathsf{H}$, we explore the possibility of imposing $\delta^{(2)} = 0$. This condition can indeed be satisfied by an appropriate choice of the $C_i$, which also serves to clarify their role as field redefinition ambiguities. After this choice, the quartic Lagrangian takes the form:

{\small
\begin{widetext}
\begin{align}
\mathcal{L}^{(4)} = \;
g^2 \Bigl\{
&M^2 \Bigl[
- \frac{1}{8}\,H_{bc} H^{bc} V_a V^a
+ \frac{1}{16}\,H^b{}_b H^c{}_c V_a V^a
+ \frac{1}{2}\,H^c{}_a H_{bc} V^a V^b
- \frac{1}{4}\,H_{ab} H^c{}_c V^a V^b
\Bigr] \notag \\[1pt]
&+ M \Bigl[
- \frac{1}{2}\,H^d{}_b H_{bc} V^a\,\partial_a H^c{}_d
+ \frac{1}{4}\,H^b{}_b H^{cd} V^a\,\partial_a H_{cd}
+ \frac{1}{8}\,H_{bc} H^{bc} V^a\,\partial_a H^d{}_d
- \frac{1}{16}\,H^b{}_b H^c{}_c V^a\,\partial_a H^d{}_d \notag \\
&\phantom{+M\Bigl[}
- \frac{1}{2}\,H^b{}_a H^{cd} V^a\,\partial_b H_{cd}
+ \frac{1}{4}\,H^b{}_a H^c{}_c V^a\,\partial_b H^d{}_d
- \frac{1}{2}\,H^b{}_a H_{bc} V^a\,\partial^c H^d{}_d
\Bigr] \notag \\[1pt]
&- \frac{1}{8}\,H_{cd} H^{cd}\,\partial_a V_b\,\partial^b V^a
+ \frac{1}{16}\,H^c{}_c H^d{}_d\,\partial_a V_b\,\partial^b V^a
+ \frac{1}{8}\,H_{cd} H^{cd}\,\partial_b V_a\,\partial^b V^a
- \frac{1}{16}\,H^c{}_c H^d{}_d\,\partial_b V_a\,\partial^b V^a \notag \\
&- \frac{1}{2}\,H^d{}_a H_{cd}\,\partial_b V^c\,\partial^b V^a
+ \frac{1}{4}\,H_{ac} H^d{}_d\,\partial_b V^c\,\partial^b V^a
- \frac{1}{2}\,H^d{}_b H_{cd}\,\partial^b V^a\,\partial^c V_a
+ \frac{1}{4}\,H_{bc} H^d{}_d\,\partial^b V^a\,\partial^c V_a \notag \\
&+ H^d{}_b H_{cd}\,\partial_a V^a\,\partial^c V^b
- \frac{1}{2}\,H_{bc} H^d{}_d\,\partial_a V^a\,\partial^c V^b
+ \frac{1}{2}\,H_{ad} H_{bc}\,\partial^b V^a\,\partial^d V^c
- \frac{1}{2}\,H_{ac} H_{bd}\,\partial^b V^a\,\partial^d V^c \notag\\[1pt]
&- \frac{1}{2}\,V^a V^b\,\partial_b H^d{}_d\,\partial_c H_a{}^c
- \frac{1}{2}\,V^a V^b\,\partial_b H_{ac}\,\partial^c H^d{}_d
- \frac{1}{2}\,H^c{}_a V^a\,\partial_b V^b\,\partial_c H^d{}_d \notag \\
&- \frac{1}{2}\,H^d{}_d V^a\,\partial_a H_{bc}\,\partial^c V^b
- \frac{1}{2}\,H_{bc} V^a\,\partial_a H^d{}_d\,\partial^c V^b
- H^d{}_c V^a\,\partial_b H_{ad}\,\partial^c V^b
- H^d{}_a V^a\,\partial_b H_{cd}\,\partial^c V^b \notag \\
&+ H^{cd} V^a\,\partial_b V^b\,\partial_d H_{ac}
- \frac{1}{2}\,H^c{}_c V^a\,\partial_b V^b\,\partial_d H_a{}^d
+ H^c{}_a V^a\,\partial_b V^b\,\partial_d H_c{}^d \notag\\[1pt]
&- \frac{1}{2}\,H^c{}_a V^a V^b\,\partial_c\partial_b H^d{}_d
- \frac{1}{2}\,H^c{}_c V^a V^b\,\partial_d\partial_b H_a{}^d \notag\\[1pt]
&- \frac{1}{2}\,H^{ab} H^{cd}\,\partial_b H_{de}\,\partial_c H_a{}^e
+ \frac{1}{4}\,H^c{}_a H^{ab}\,\partial_b H^{de}\,\partial_c H_{de}
- \frac{1}{8}\,H^a{}_a H^{bc}\,\partial_b H^{de}\,\partial_c H_{de}
- \frac{1}{8}\,H^c{}_a H^{ab}\,\partial_b H^d{}_d\,\partial_c H^e{}_e\notag \\
&
+ \frac{1}{16}\,H^a{}_a H^{bc}\,\partial_b H^d{}_d\,\partial_c H^e{}_e + \frac{1}{2}\,H^{ab} H^{cd}\,\partial_c H_{ae}\,\partial_d H_b{}^e
+ \frac{1}{2}\,H^{ab} H^{cd}\,\partial_b H_{ac}\,\partial_d H^e{}_e
- \frac{1}{4}\,H^{ab} H^{cd}\,\partial_c H_{ab}\,\partial_d H^e{}_e \notag \\
&- \frac{1}{4}\,H^c{}_a H^{ab}\,\partial_d H^e{}_e\,\partial^d H_{bc}
+ \frac{1}{8}\,H^a{}_a H^{bc}\,\partial_d H^e{}_e\,\partial^d H_{bc}
+ \frac{1}{32}\,H_{ab} H^{ab}\,\partial_d H^e{}_e\,\partial^d H^c{}_c
- \frac{1}{64}\,H^a{}_a H^b{}_b\,\partial_d H^e{}_e\,\partial^d H^c{}_c \notag\\[1pt]
&+ H^{ab} H^{cd}\,\partial_d H_{ce}\,\partial^e H_{ab}
- \frac{1}{8}\,H^{ab} H^{cd}\,\partial_e H_{cd}\,\partial^e H_{ab}
- H^{ab} H^{cd}\,\partial_d H_{be}\,\partial^e H_{ac}
+ \frac{1}{4}\,H^{ab} H^{cd}\,\partial_e H_{bd}\,\partial^e H_{ac} \notag\\[1pt]
&+ \frac{1}{2}\,H^a{}_a H^{bc}\,\partial_d H_b{}^d\,\partial_e H_c{}^e
+ \frac{1}{2}\,H^a{}_a H^{bc}\,\partial_c H_b{}^d\,\partial_e H_d{}^e
- \frac{1}{4}\,H_{ab} H^{ab}\,\partial_c H^{cd}\,\partial_e H_d{}^e
- \frac{1}{8}\,H^a{}_a H^b{}_b\,\partial_c H^{cd}\,\partial_e H_d{}^e \notag \\
&+ H^c{}_a H^{ab}\,\partial^d H_{bc}\,\partial_e H_d{}^e
- \frac{1}{2}\,H^a{}_a H^{bc}\,\partial^d H_{bc}\,\partial_e H_d{}^e
- \frac{1}{4}\,H_{ab} H^{ab}\,\partial^d H^c{}_c\,\partial_e H_d{}^e
+ \frac{1}{8}\,H^a{}_a H^b{}_b\,\partial^d H^c{}_c\,\partial_e H_d{}^e \notag\\[1pt]
&- H^c{}_a H^{ab}\,\partial_c H_{de}\,\partial^e H_b{}^d
- \frac{1}{2}\,H^c{}_a H^{ab}\,\partial_d H_{ce}\,\partial^e H_b{}^d
- \frac{1}{4}\,H^a{}_a H^{bc}\,\partial_d H_{ce}\,\partial^e H_b{}^d
+ \frac{1}{2}\,H^c{}_a H^{ab}\,\partial_e H_{cd}\,\partial^e H_b{}^d \notag \\
&- \frac{1}{4}\,H^a{}_a H^{bc}\,\partial_e H_{cd}\,\partial^e H_b{}^d 
+ \frac{3}{8}\,H_{ab} H^{ab}\,\partial_d H_{ce}\,\partial^e H^{cd}
+ \frac{1}{16}\,H^a{}_a H^b{}_b\,\partial_d H_{ce}\,\partial^e H^{cd}
- \frac{1}{16}\,H_{ab} H^{ab}\,\partial_e H_{cd}\,\partial^e H^{cd} \notag \\
& + \frac{1}{32}\,H^a{}_a H^b{}_b\,\partial_e H_{cd}\,\partial^e H^{cd}
\Bigr\}
\end{align}
\end{widetext}
}

Again, we encounter power-counting renormalizable operators parametrically interlaced with higher-order ones. 
\section{Comments and conclusion}
Several points emerge from our analysis. Some are straightforward observations concerning the mathematical output, others are more speculative and touch on physical interpretations. We present them as a list, and hope that the more tentative remarks, admittedly closer to conjectures than conclusions, do not overshadow the more solid ones.

\begin{itemize}
    \item A gauge-symmetric model exists that propagates a massless spin-2 and a massive spin-1 particle Eq.~(\ref{lagVH}), with the trace of $\mathsf{H}$ acting as a Goldstone boson. 

    \item The mixed $\mathsf{V}$-$\mathsf{H}$ system  admits a non-trivial consistent deformation that survives the  $\mathcal{O}(g^2)$ consistency requirements of the Noether procedure.
    

  \item  Whether the propagating spin-2 state can be fully identified with the graviton remains an open question. The structure of the obtained couplings is suggestive of this identification, but the prominent role of the trace may introduce phenomenological complications. A complete analysis incorporating the effects of the interactions on both $\mathsf{H}$ and $\mathsf{V}$ is required before any definitive conclusion can be drawn.

 \item The decoupling of a massive mixed state is non-trivial. We intend to investigate whether the model obtained upon integrating out the heavy vector is fully equivalent to GR.

 \item The presence of a dynamical vector field intertwined with the long-range propagation of a massless spin-2 state opens the possibility of new indirect gravitational couplings, with the vector field serving as a portal, an appealing feature given the relative ease with which vector fields couple to matter.

  \item It is abundantly known that the long-range gravitational behavior demands dominant contributions from dark-energy and dark-matter. It is tempting to speculate whether such effects could be connected to the exotic interactions met allowing mixed representations involving vector fields and other matter content. 

  \item Previous attempts to construct multicomponent graviton models encountered obstructions and loss of unitarity \cite{Boulanger:2000rq}. As demonstrated here, distinct models with the same field content but from a different kinetic structure, can be obtained. It would be of considerable interest to investigate multicomponent self-interactions of Yang-Mills type, using these novel kinetic starting points as a foundation. 

  \item \textit{Towards covariantization.} The recognizable Lie-derivative form of the vector gauge transformation in Eq.~(\ref{surprise}) is particularly suggestive. The full transformation,
    \begin{align}
        &\delta V_m = M^{-1}\partial_m (\partial \cdot \xi) + g (\xi^a \partial_a V_m
        + V^a \partial_m \xi_a ) \, , \notag 
    \end{align}
    mirrors the transformation law of $\Gamma{}_{m}$, the trace of the affine connection $\Gamma^l{}_{mn}$. This structural correspondence hints at an achievable covariantization of our bottom-up structures, deserving further scrutiny. 

\item One may be tempted to interpret much of the presented model as a Stückelberg mechanism applied to Proca theory, with the trace of $\mathsf{H}$ playing the role of the compensating scalar rather than an independent field. However, since the Stückelberg procedure is not a field redefinition but rather a formal substitution, applying it to a consistent model of interacting spin-1 and spin-2 fields does not guarantee the consistency of the full non-linear theory. The bottom-up Noether procedure adopted here makes this point unambiguous.


\end{itemize}

\begin{acknowledgments}
I am grateful to Will Barker, Dario Francia and Alessandro Santoni for illuminating discussions. 
I gratefully acknowledge the hospitality of the Department of Physics at the University of Calabria during my visit.
This work was supported by the Estonian Research Council grant PRG1677 and the CoE program TK202 'Fundamental Universe'. 
\end{acknowledgments}

\appendix

\section{Notation for index contractions} \label{notation}
To handle the proliferation of independent index contractions and equivalences arising from integration by parts (IBP), we employ a compact symbolic notation in which the fundamental fields are represented by index-free \emph{sans-serif} symbols,
\begin{align}
 \mathsf{H} \equiv H_{ab}, \qquad \mathsf{V} \equiv V_{a}\,, 
\qquad  \mathsf{\xi}  \equiv \xi_{a} \, .
\end{align}
A monomial symbol such as $\mathsf{H}^2 \partial \mathsf{V}$ denotes the complete set of inequivalent contractions of the corresponding fields and derivatives,
\begin{align} \label{app:symb}
\mathsf{H}^2 \partial \mathsf{V} \equiv \; 
& c_1 H_{ab} \partial^{c} H_{c}{}^{b} V^{a} +
  c_2 H_{ab} \partial^{c} H_{c}{}^{a} V^{b} \nonumber \\
+ \; & c_3 H_{ab} \partial^{b} H^{a}{}_{c} V^{c}  
     + c_4 H^{ab} \partial_{a} H_{bc} V^{c} + \cdots
\end{align}
The ordering of symbols and the placement of the derivative operator carry no meaning, so that $\mathsf{H}^2 \partial \mathsf{V}$, 
$\partial \mathsf{H}^2 \mathsf{V}$, and $\mathsf{V} \partial \mathsf{H}^2$ all refer to the same set of terms in Eq.~\eqref{app:symb}. It is understood that a canonicalization of terms is used to avoid double counting due to IBP. Moreover, the equivalences of tensorial objects are always reached via a functional derivation which automatically nullifies boundary terms.

\bibliography{sample}

\begin{thebibliography}{37}%
\makeatletter
\providecommand \@ifxundefined [1]{%
 \@ifx{#1\undefined}
}%
\providecommand \@ifnum [1]{%
 \ifnum #1\expandafter \@firstoftwo
 \else \expandafter \@secondoftwo
 \fi
}%
\providecommand \@ifx [1]{%
 \ifx #1\expandafter \@firstoftwo
 \else \expandafter \@secondoftwo
 \fi
}%
\providecommand \natexlab [1]{#1}%
\providecommand \enquote  [1]{``#1''}%
\providecommand \bibnamefont  [1]{#1}%
\providecommand \bibfnamefont [1]{#1}%
\providecommand \citenamefont [1]{#1}%
\providecommand \href@noop [0]{\@secondoftwo}%
\providecommand \href [0]{\begingroup \@sanitize@url \@href}%
\providecommand \@href[1]{\@@startlink{#1}\@@href}%
\providecommand \@@href[1]{\endgroup#1\@@endlink}%
\providecommand \@sanitize@url [0]{\catcode `\\12\catcode `\$12\catcode `\&12\catcode `\#12\catcode `\^12\catcode `\_12\catcode `\%12\relax}%
\providecommand \@@startlink[1]{}%
\providecommand \@@endlink[0]{}%
\providecommand \url  [0]{\begingroup\@sanitize@url \@url }%
\providecommand \@url [1]{\endgroup\@href {#1}{\urlprefix }}%
\providecommand \urlprefix  [0]{URL }%
\providecommand \Eprint [0]{\href }%
\providecommand \doibase [0]{https://doi.org/}%
\providecommand \selectlanguage [0]{\@gobble}%
\providecommand \bibinfo  [0]{\@secondoftwo}%
\providecommand \bibfield  [0]{\@secondoftwo}%
\providecommand \translation [1]{[#1]}%
\providecommand \BibitemOpen [0]{}%
\providecommand \bibitemStop [0]{}%
\providecommand \bibitemNoStop [0]{.\EOS\space}%
\providecommand \EOS [0]{\spacefactor3000\relax}%
\providecommand \BibitemShut  [1]{\csname bibitem#1\endcsname}%
\let\auto@bib@innerbib\@empty
\bibitem [{\citenamefont {Gomis}\ and\ \citenamefont {Weinberg}(1996)}]{Gomis:1995jp}%
  \BibitemOpen
  \bibfield  {author} {\bibinfo {author} {\bibfnamefont {J.}~\bibnamefont {Gomis}}\ and\ \bibinfo {author} {\bibfnamefont {S.}~\bibnamefont {Weinberg}},\ }\bibfield  {title} {\bibinfo {title} {{Are nonrenormalizable gauge theories renormalizable?}},\ }\href {https://doi.org/10.1016/0550-3213(96)00132-0} {\bibfield  {journal} {\bibinfo  {journal} {Nucl. Phys. B}\ }\textbf {\bibinfo {volume} {469}},\ \bibinfo {pages} {473} (\bibinfo {year} {1996})},\ \Eprint {https://arxiv.org/abs/hep-th/9510087} {arXiv:hep-th/9510087} \BibitemShut {NoStop}%
\bibitem [{\citenamefont {Donoghue}(1994{\natexlab{a}})}]{Donoghue:1993eb}%
  \BibitemOpen
  \bibfield  {author} {\bibinfo {author} {\bibfnamefont {J.~F.}\ \bibnamefont {Donoghue}},\ }\bibfield  {title} {\bibinfo {title} {{Leading quantum correction to the Newtonian potential}},\ }\href {https://doi.org/10.1103/PhysRevLett.72.2996} {\bibfield  {journal} {\bibinfo  {journal} {Phys. Rev. Lett.}\ }\textbf {\bibinfo {volume} {72}},\ \bibinfo {pages} {2996} (\bibinfo {year} {1994}{\natexlab{a}})},\ \Eprint {https://arxiv.org/abs/gr-qc/9310024} {arXiv:gr-qc/9310024} \BibitemShut {NoStop}%
\bibitem [{\citenamefont {Donoghue}(1994{\natexlab{b}})}]{Donoghue:1994dn}%
  \BibitemOpen
  \bibfield  {author} {\bibinfo {author} {\bibfnamefont {J.~F.}\ \bibnamefont {Donoghue}},\ }\bibfield  {title} {\bibinfo {title} {{General relativity as an effective field theory: The leading quantum corrections}},\ }\href {https://doi.org/10.1103/PhysRevD.50.3874} {\bibfield  {journal} {\bibinfo  {journal} {Phys. Rev. D}\ }\textbf {\bibinfo {volume} {50}},\ \bibinfo {pages} {3874} (\bibinfo {year} {1994}{\natexlab{b}})},\ \Eprint {https://arxiv.org/abs/gr-qc/9405057} {arXiv:gr-qc/9405057} \BibitemShut {NoStop}%
\bibitem [{\citenamefont {Gasser}\ and\ \citenamefont {Leutwyler}(1984)}]{Gasser:1983yg}%
  \BibitemOpen
  \bibfield  {author} {\bibinfo {author} {\bibfnamefont {J.}~\bibnamefont {Gasser}}\ and\ \bibinfo {author} {\bibfnamefont {H.}~\bibnamefont {Leutwyler}},\ }\bibfield  {title} {\bibinfo {title} {{Chiral Perturbation Theory to One Loop}},\ }\href {https://doi.org/10.1016/0003-4916(84)90242-2} {\bibfield  {journal} {\bibinfo  {journal} {Annals Phys.}\ }\textbf {\bibinfo {volume} {158}},\ \bibinfo {pages} {142} (\bibinfo {year} {1984})}\BibitemShut {NoStop}%
\bibitem [{\citenamefont {Entem}\ and\ \citenamefont {Machleidt}(2003)}]{Entem:2003ft}%
  \BibitemOpen
  \bibfield  {author} {\bibinfo {author} {\bibfnamefont {D.~R.}\ \bibnamefont {Entem}}\ and\ \bibinfo {author} {\bibfnamefont {R.}~\bibnamefont {Machleidt}},\ }\bibfield  {title} {\bibinfo {title} {{Accurate charge dependent nucleon nucleon potential at fourth order of chiral perturbation theory}},\ }\href {https://doi.org/10.1103/PhysRevC.68.041001} {\bibfield  {journal} {\bibinfo  {journal} {Phys. Rev. C}\ }\textbf {\bibinfo {volume} {68}},\ \bibinfo {pages} {041001} (\bibinfo {year} {2003})},\ \Eprint {https://arxiv.org/abs/nucl-th/0304018} {arXiv:nucl-th/0304018} \BibitemShut {NoStop}%
\bibitem [{\citenamefont {Marzo}(2022)}]{Marzo:2021iok}%
  \BibitemOpen
  \bibfield  {author} {\bibinfo {author} {\bibfnamefont {C.}~\bibnamefont {Marzo}},\ }\bibfield  {title} {\bibinfo {title} {{Radiatively stable ghost and tachyon freedom in metric affine gravity}},\ }\href {https://doi.org/10.1103/PhysRevD.106.024045} {\bibfield  {journal} {\bibinfo  {journal} {Phys. Rev. D}\ }\textbf {\bibinfo {volume} {106}},\ \bibinfo {pages} {024045} (\bibinfo {year} {2022})},\ \Eprint {https://arxiv.org/abs/2110.14788} {arXiv:2110.14788 [hep-th]} \BibitemShut {NoStop}%
\bibitem [{\citenamefont {Marzo}(2025)}]{Marzo:2024pyn}%
  \BibitemOpen
  \bibfield  {author} {\bibinfo {author} {\bibfnamefont {C.}~\bibnamefont {Marzo}},\ }\bibfield  {title} {\bibinfo {title} {{Can MAG be a predictive EFT? Radiative stability and ghost resurgence in massive vector models}},\ }\href {https://doi.org/10.1088/1361-6382/adc9f1} {\bibfield  {journal} {\bibinfo  {journal} {Class. Quant. Grav.}\ }\textbf {\bibinfo {volume} {42}},\ \bibinfo {pages} {095007} (\bibinfo {year} {2025})},\ \Eprint {https://arxiv.org/abs/2403.15003} {arXiv:2403.15003 [hep-th]} \BibitemShut {NoStop}%
\bibitem [{\citenamefont {Barker}\ \emph {et~al.}(2025{\natexlab{a}})\citenamefont {Barker}, \citenamefont {Marzo},\ and\ \citenamefont {Santoni}}]{Barker:2025xzd}%
  \BibitemOpen
  \bibfield  {author} {\bibinfo {author} {\bibfnamefont {W.}~\bibnamefont {Barker}}, \bibinfo {author} {\bibfnamefont {C.}~\bibnamefont {Marzo}},\ and\ \bibinfo {author} {\bibfnamefont {A.}~\bibnamefont {Santoni}},\ }\bibfield  {title} {\bibinfo {title} {{Can metric-affine gravity be saved?}},\ }\href {https://doi.org/10.1103/1xqx-f57g} {\bibfield  {journal} {\bibinfo  {journal} {Phys. Rev. D}\ }\textbf {\bibinfo {volume} {112}},\ \bibinfo {pages} {044032} (\bibinfo {year} {2025}{\natexlab{a}})},\ \Eprint {https://arxiv.org/abs/2505.23894} {arXiv:2505.23894 [hep-th]} \BibitemShut {NoStop}%
\bibitem [{\citenamefont {Barker}\ \emph {et~al.}(2025{\natexlab{b}})\citenamefont {Barker}, \citenamefont {Marzo},\ and\ \citenamefont {Santoni}}]{Barker:2025rzd}%
  \BibitemOpen
  \bibfield  {author} {\bibinfo {author} {\bibfnamefont {W.}~\bibnamefont {Barker}}, \bibinfo {author} {\bibfnamefont {C.}~\bibnamefont {Marzo}},\ and\ \bibinfo {author} {\bibfnamefont {A.}~\bibnamefont {Santoni}},\ }\bibfield  {title} {\bibinfo {title} {{Infrared foundations for quantum geometry I: Catalogue of totally symmetric rank-three field theories}},\ }\href@noop {} {\  (\bibinfo {year} {2025}{\natexlab{b}})},\ \Eprint {https://arxiv.org/abs/2506.21662} {arXiv:2506.21662 [hep-th]} \BibitemShut {NoStop}%
\bibitem [{\citenamefont {Barker}\ \emph {et~al.}(2025{\natexlab{c}})\citenamefont {Barker}, \citenamefont {Marzo},\ and\ \citenamefont {Santoni}}]{Barker:2025fgo}%
  \BibitemOpen
  \bibfield  {author} {\bibinfo {author} {\bibfnamefont {W.}~\bibnamefont {Barker}}, \bibinfo {author} {\bibfnamefont {C.}~\bibnamefont {Marzo}},\ and\ \bibinfo {author} {\bibfnamefont {A.}~\bibnamefont {Santoni}},\ }\bibfield  {title} {\bibinfo {title} {{Infrared foundations for quantum geometry II: Catalogue of all torsion-like theories including new ghost-tachyon-free cases}},\ }\href@noop {} {\  (\bibinfo {year} {2025}{\natexlab{c}})},\ \Eprint {https://arxiv.org/abs/2507.05349} {arXiv:2507.05349 [hep-th]} \BibitemShut {NoStop}%
\bibitem [{\citenamefont {Fang}\ and\ \citenamefont {Fronsdal}(1979)}]{Fang:1978rc}%
  \BibitemOpen
  \bibfield  {author} {\bibinfo {author} {\bibfnamefont {J.}~\bibnamefont {Fang}}\ and\ \bibinfo {author} {\bibfnamefont {C.}~\bibnamefont {Fronsdal}},\ }\bibfield  {title} {\bibinfo {title} {{Deformation of Gauge Groups. Gravitation}},\ }\href {https://doi.org/10.1063/1.524007} {\bibfield  {journal} {\bibinfo  {journal} {J. Math. Phys.}\ }\textbf {\bibinfo {volume} {20}},\ \bibinfo {pages} {2264} (\bibinfo {year} {1979})}\BibitemShut {NoStop}%
\bibitem [{\citenamefont {Deser}\ and\ \citenamefont {Arnowitt}(1963)}]{Deser:1963zzc}%
  \BibitemOpen
  \bibfield  {author} {\bibinfo {author} {\bibfnamefont {S.}~\bibnamefont {Deser}}\ and\ \bibinfo {author} {\bibfnamefont {R.}~\bibnamefont {Arnowitt}},\ }\bibfield  {title} {\bibinfo {title} {{Interaction Among Gauge Vector Fields}},\ }\href {https://doi.org/10.1016/0029-5582(63)90081-6} {\bibfield  {journal} {\bibinfo  {journal} {Nucl. Phys.}\ }\textbf {\bibinfo {volume} {49}},\ \bibinfo {pages} {133} (\bibinfo {year} {1963})}\BibitemShut {NoStop}%
\bibitem [{\citenamefont {Berends}\ \emph {et~al.}(1985)\citenamefont {Berends}, \citenamefont {Burgers},\ and\ \citenamefont {van Dam}}]{Berends:1984rq}%
  \BibitemOpen
  \bibfield  {author} {\bibinfo {author} {\bibfnamefont {F.~A.}\ \bibnamefont {Berends}}, \bibinfo {author} {\bibfnamefont {G.~J.~H.}\ \bibnamefont {Burgers}},\ and\ \bibinfo {author} {\bibfnamefont {H.}~\bibnamefont {van Dam}},\ }\bibfield  {title} {\bibinfo {title} {{On the Theoretical Problems in Constructing Interactions Involving Higher Spin Massless Particles}},\ }\href {https://doi.org/10.1016/0550-3213(85)90074-4} {\bibfield  {journal} {\bibinfo  {journal} {Nucl. Phys. B}\ }\textbf {\bibinfo {volume} {260}},\ \bibinfo {pages} {295} (\bibinfo {year} {1985})}\BibitemShut {NoStop}%
\bibitem [{\citenamefont {Boulanger}\ \emph {et~al.}(2001)\citenamefont {Boulanger}, \citenamefont {Damour}, \citenamefont {Gualtieri},\ and\ \citenamefont {Henneaux}}]{Boulanger:2000rq}%
  \BibitemOpen
  \bibfield  {author} {\bibinfo {author} {\bibfnamefont {N.}~\bibnamefont {Boulanger}}, \bibinfo {author} {\bibfnamefont {T.}~\bibnamefont {Damour}}, \bibinfo {author} {\bibfnamefont {L.}~\bibnamefont {Gualtieri}},\ and\ \bibinfo {author} {\bibfnamefont {M.}~\bibnamefont {Henneaux}},\ }\bibfield  {title} {\bibinfo {title} {{Inconsistency of interacting, multigraviton theories}},\ }\href {https://doi.org/10.1016/S0550-3213(00)00718-5} {\bibfield  {journal} {\bibinfo  {journal} {Nucl. Phys. B}\ }\textbf {\bibinfo {volume} {597}},\ \bibinfo {pages} {127} (\bibinfo {year} {2001})},\ \Eprint {https://arxiv.org/abs/hep-th/0007220} {arXiv:hep-th/0007220} \BibitemShut {NoStop}%
\bibitem [{\citenamefont {Barnich}\ \emph {et~al.}(2018)\citenamefont {Barnich}, \citenamefont {Boulanger}, \citenamefont {Henneaux}, \citenamefont {Julia}, \citenamefont {Lekeu},\ and\ \citenamefont {Ranjbar}}]{Barnich:2017nty}%
  \BibitemOpen
  \bibfield  {author} {\bibinfo {author} {\bibfnamefont {G.}~\bibnamefont {Barnich}}, \bibinfo {author} {\bibfnamefont {N.}~\bibnamefont {Boulanger}}, \bibinfo {author} {\bibfnamefont {M.}~\bibnamefont {Henneaux}}, \bibinfo {author} {\bibfnamefont {B.}~\bibnamefont {Julia}}, \bibinfo {author} {\bibfnamefont {V.}~\bibnamefont {Lekeu}},\ and\ \bibinfo {author} {\bibfnamefont {A.}~\bibnamefont {Ranjbar}},\ }\bibfield  {title} {\bibinfo {title} {{Deformations of vector-scalar models}},\ }\href {https://doi.org/10.1007/JHEP02(2018)064} {\bibfield  {journal} {\bibinfo  {journal} {JHEP}\ }\textbf {\bibinfo {volume} {02}},\ \bibinfo {pages} {064}},\ \Eprint {https://arxiv.org/abs/1712.08126} {arXiv:1712.08126 [hep-th]} \BibitemShut {NoStop}%
\bibitem [{\citenamefont {Zinov'ev}(2017)}]{Zinovev:2017cdp}%
  \BibitemOpen
  \bibfield  {author} {\bibinfo {author} {\bibfnamefont {Y.~M.}\ \bibnamefont {Zinov'ev}},\ }\bibfield  {title} {\bibinfo {title} {{Gravity from the constructive approach standpoint}},\ }\href {https://doi.org/10.1134/S0040577917050051} {\bibfield  {journal} {\bibinfo  {journal} {Theor. Math. Phys.}\ }\textbf {\bibinfo {volume} {191}},\ \bibinfo {pages} {655} (\bibinfo {year} {2017})}\BibitemShut {NoStop}%
\bibitem [{\citenamefont {Wald}(1986)}]{Wald:1986bj}%
  \BibitemOpen
  \bibfield  {author} {\bibinfo {author} {\bibfnamefont {R.~M.}\ \bibnamefont {Wald}},\ }\bibfield  {title} {\bibinfo {title} {{Spin-2 Fields and General Covariance}},\ }\href {https://doi.org/10.1103/PhysRevD.33.3613} {\bibfield  {journal} {\bibinfo  {journal} {Phys. Rev. D}\ }\textbf {\bibinfo {volume} {33}},\ \bibinfo {pages} {3613} (\bibinfo {year} {1986})}\BibitemShut {NoStop}%
\bibitem [{\citenamefont {Cutler}\ and\ \citenamefont {Wald}(1987)}]{Cutler:1986dv}%
  \BibitemOpen
  \bibfield  {author} {\bibinfo {author} {\bibfnamefont {C.}~\bibnamefont {Cutler}}\ and\ \bibinfo {author} {\bibfnamefont {R.~M.}\ \bibnamefont {Wald}},\ }\bibfield  {title} {\bibinfo {title} {{A New Type of Gauge Invariance for a Collection of Massless Spin-2 Fields. 1. Existence and Uniqueness}},\ }\href {https://doi.org/10.1088/0264-9381/4/5/024} {\bibfield  {journal} {\bibinfo  {journal} {Class. Quant. Grav.}\ }\textbf {\bibinfo {volume} {4}},\ \bibinfo {pages} {1267} (\bibinfo {year} {1987})}\BibitemShut {NoStop}%
\bibitem [{\citenamefont {Fierz}\ and\ \citenamefont {Pauli}(1939)}]{Fierz:1939ix}%
  \BibitemOpen
  \bibfield  {author} {\bibinfo {author} {\bibfnamefont {M.}~\bibnamefont {Fierz}}\ and\ \bibinfo {author} {\bibfnamefont {W.}~\bibnamefont {Pauli}},\ }\bibfield  {title} {\bibinfo {title} {{On relativistic wave equations for particles of arbitrary spin in an electromagnetic field}},\ }\href {https://doi.org/10.1098/rspa.1939.0140} {\bibfield  {journal} {\bibinfo  {journal} {Proc. Roy. Soc. Lond. A}\ }\textbf {\bibinfo {volume} {173}},\ \bibinfo {pages} {211} (\bibinfo {year} {1939})}\BibitemShut {NoStop}%
\bibitem [{\citenamefont {Percacci}\ and\ \citenamefont {Sezgin}(2020)}]{Percacci:2020ddy}%
  \BibitemOpen
  \bibfield  {author} {\bibinfo {author} {\bibfnamefont {R.}~\bibnamefont {Percacci}}\ and\ \bibinfo {author} {\bibfnamefont {E.}~\bibnamefont {Sezgin}},\ }\bibfield  {title} {\bibinfo {title} {{New class of ghost- and tachyon-free metric affine gravities}},\ }\href {https://doi.org/10.1103/PhysRevD.101.084040} {\bibfield  {journal} {\bibinfo  {journal} {Phys. Rev. D}\ }\textbf {\bibinfo {volume} {101}},\ \bibinfo {pages} {084040} (\bibinfo {year} {2020})},\ \bibinfo {note} {[Erratum: Phys.Rev.D 111, 109902 (2025)]},\ \Eprint {https://arxiv.org/abs/1912.01023} {arXiv:1912.01023 [hep-th]} \BibitemShut {NoStop}%
\bibitem [{\citenamefont {van~der Bij}\ \emph {et~al.}(1982)\citenamefont {van~der Bij}, \citenamefont {van Dam},\ and\ \citenamefont {Ng}}]{vanderBij:1981ym}%
  \BibitemOpen
  \bibfield  {author} {\bibinfo {author} {\bibfnamefont {J.~J.}\ \bibnamefont {van~der Bij}}, \bibinfo {author} {\bibfnamefont {H.}~\bibnamefont {van Dam}},\ and\ \bibinfo {author} {\bibfnamefont {Y.~J.}\ \bibnamefont {Ng}},\ }\bibfield  {title} {\bibinfo {title} {{The Exchange of Massless Spin Two Particles}},\ }\href {https://doi.org/10.1016/0378-4371(82)90247-3} {\bibfield  {journal} {\bibinfo  {journal} {Physica A}\ }\textbf {\bibinfo {volume} {116}},\ \bibinfo {pages} {307} (\bibinfo {year} {1982})}\BibitemShut {NoStop}%
\bibitem [{\citenamefont {Barker}\ \emph {et~al.}(2025{\natexlab{d}})\citenamefont {Barker}, \citenamefont {Marzo},\ and\ \citenamefont {Rigouzzo}}]{Barker:2024juc}%
  \BibitemOpen
  \bibfield  {author} {\bibinfo {author} {\bibfnamefont {W.}~\bibnamefont {Barker}}, \bibinfo {author} {\bibfnamefont {C.}~\bibnamefont {Marzo}},\ and\ \bibinfo {author} {\bibfnamefont {C.}~\bibnamefont {Rigouzzo}},\ }\bibfield  {title} {\bibinfo {title} {{Particle spectrum for any tensor Lagrangian}},\ }\href {https://doi.org/10.1103/PhysRevD.112.016018} {\bibfield  {journal} {\bibinfo  {journal} {Phys. Rev. D}\ }\textbf {\bibinfo {volume} {112}},\ \bibinfo {pages} {016018} (\bibinfo {year} {2025}{\natexlab{d}})},\ \Eprint {https://arxiv.org/abs/2406.09500} {arXiv:2406.09500 [hep-th]} \BibitemShut {NoStop}%
\bibitem [{\citenamefont {Henneaux}(1990)}]{Henneaux:1989jq}%
  \BibitemOpen
  \bibfield  {author} {\bibinfo {author} {\bibfnamefont {M.}~\bibnamefont {Henneaux}},\ }\bibfield  {title} {\bibinfo {title} {{Lectures on the Antifield-BRST Formalism for Gauge Theories}},\ }\href {https://doi.org/10.1016/0920-5632(90)90647-D} {\bibfield  {journal} {\bibinfo  {journal} {Nucl. Phys. B Proc. Suppl.}\ }\textbf {\bibinfo {volume} {18}},\ \bibinfo {pages} {47} (\bibinfo {year} {1990})}\BibitemShut {NoStop}%
\bibitem [{\citenamefont {Barnich}\ \emph {et~al.}(2026)\citenamefont {Barnich}, \citenamefont {Baulieu}, \citenamefont {Henneaux},\ and\ \citenamefont {Wetzstein}}]{Barnich:2025jna}%
  \BibitemOpen
  \bibfield  {author} {\bibinfo {author} {\bibfnamefont {G.}~\bibnamefont {Barnich}}, \bibinfo {author} {\bibfnamefont {L.}~\bibnamefont {Baulieu}}, \bibinfo {author} {\bibfnamefont {M.}~\bibnamefont {Henneaux}},\ and\ \bibinfo {author} {\bibfnamefont {T.}~\bibnamefont {Wetzstein}},\ }\bibfield  {title} {\bibinfo {title} {{BV-BRST Noether theorem}},\ }\href {https://doi.org/10.1007/JHEP02(2026)100} {\bibfield  {journal} {\bibinfo  {journal} {JHEP}\ }\textbf {\bibinfo {volume} {02}},\ \bibinfo {pages} {100}},\ \Eprint {https://arxiv.org/abs/2512.19418} {arXiv:2512.19418 [hep-th]} \BibitemShut {NoStop}%
\bibitem [{\citenamefont {Barnich}\ \emph {et~al.}(2000)\citenamefont {Barnich}, \citenamefont {Brandt},\ and\ \citenamefont {Henneaux}}]{Barnich:2000zw}%
  \BibitemOpen
  \bibfield  {author} {\bibinfo {author} {\bibfnamefont {G.}~\bibnamefont {Barnich}}, \bibinfo {author} {\bibfnamefont {F.}~\bibnamefont {Brandt}},\ and\ \bibinfo {author} {\bibfnamefont {M.}~\bibnamefont {Henneaux}},\ }\bibfield  {title} {\bibinfo {title} {{Local BRST cohomology in gauge theories}},\ }\href {https://doi.org/10.1016/S0370-1573(00)00049-1} {\bibfield  {journal} {\bibinfo  {journal} {Phys. Rept.}\ }\textbf {\bibinfo {volume} {338}},\ \bibinfo {pages} {439} (\bibinfo {year} {2000})},\ \Eprint {https://arxiv.org/abs/hep-th/0002245} {arXiv:hep-th/0002245} \BibitemShut {NoStop}%
\bibitem [{\citenamefont {Henneaux}(1998)}]{Henneaux:1997bm}%
  \BibitemOpen
  \bibfield  {author} {\bibinfo {author} {\bibfnamefont {M.}~\bibnamefont {Henneaux}},\ }\bibfield  {title} {\bibinfo {title} {{Consistent interactions between gauge fields: The Cohomological approach}},\ }\href {https://doi.org/10.1090/conm/219/03070} {\bibfield  {journal} {\bibinfo  {journal} {Contemp. Math.}\ }\textbf {\bibinfo {volume} {219}},\ \bibinfo {pages} {93} (\bibinfo {year} {1998})},\ \Eprint {https://arxiv.org/abs/hep-th/9712226} {arXiv:hep-th/9712226} \BibitemShut {NoStop}%
\bibitem [{\citenamefont {Nutma}(2014)}]{Nutma:2013zea}%
  \BibitemOpen
  \bibfield  {author} {\bibinfo {author} {\bibfnamefont {T.}~\bibnamefont {Nutma}},\ }\bibfield  {title} {\bibinfo {title} {{xTras : A field-theory inspired xAct package for mathematica}},\ }\href {https://doi.org/10.1016/j.cpc.2014.02.006} {\bibfield  {journal} {\bibinfo  {journal} {Comput. Phys. Commun.}\ }\textbf {\bibinfo {volume} {185}},\ \bibinfo {pages} {1719} (\bibinfo {year} {2014})},\ \Eprint {https://arxiv.org/abs/1308.3493} {arXiv:1308.3493 [cs.SC]} \BibitemShut {NoStop}%
\bibitem [{\citenamefont {Brizuela}\ \emph {et~al.}(2009)\citenamefont {Brizuela}, \citenamefont {Martin-Garcia},\ and\ \citenamefont {Mena~Marugan}}]{Brizuela:2008ra}%
  \BibitemOpen
  \bibfield  {author} {\bibinfo {author} {\bibfnamefont {D.}~\bibnamefont {Brizuela}}, \bibinfo {author} {\bibfnamefont {J.~M.}\ \bibnamefont {Martin-Garcia}},\ and\ \bibinfo {author} {\bibfnamefont {G.~A.}\ \bibnamefont {Mena~Marugan}},\ }\bibfield  {title} {\bibinfo {title} {{xPert: Computer algebra for metric perturbation theory}},\ }\href {https://doi.org/10.1007/s10714-009-0773-2} {\bibfield  {journal} {\bibinfo  {journal} {Gen. Rel. Grav.}\ }\textbf {\bibinfo {volume} {41}},\ \bibinfo {pages} {2415} (\bibinfo {year} {2009})},\ \Eprint {https://arxiv.org/abs/0807.0824} {arXiv:0807.0824 [gr-qc]} \BibitemShut {NoStop}%
\bibitem [{\citenamefont {Kraichnan}(1955)}]{Kraichnan:1955zz}%
  \BibitemOpen
  \bibfield  {author} {\bibinfo {author} {\bibfnamefont {R.~H.}\ \bibnamefont {Kraichnan}},\ }\bibfield  {title} {\bibinfo {title} {{Special-Relativistic Derivation of Generally Covariant Gravitation Theory}},\ }\href {https://doi.org/10.1103/PhysRev.98.1118} {\bibfield  {journal} {\bibinfo  {journal} {Phys. Rev.}\ }\textbf {\bibinfo {volume} {98}},\ \bibinfo {pages} {1118} (\bibinfo {year} {1955})}\BibitemShut {NoStop}%
\bibitem [{\citenamefont {Gupta}(1954)}]{Gupta:1954zz}%
  \BibitemOpen
  \bibfield  {author} {\bibinfo {author} {\bibfnamefont {S.~N.}\ \bibnamefont {Gupta}},\ }\bibfield  {title} {\bibinfo {title} {{Gravitation and Electromagnetism}},\ }\href {https://doi.org/10.1103/PhysRev.96.1683} {\bibfield  {journal} {\bibinfo  {journal} {Phys. Rev.}\ }\textbf {\bibinfo {volume} {96}},\ \bibinfo {pages} {1683} (\bibinfo {year} {1954})}\BibitemShut {NoStop}%
\bibitem [{\citenamefont {Padmanabhan}(2008)}]{Padmanabhan:2004xk}%
  \BibitemOpen
  \bibfield  {author} {\bibinfo {author} {\bibfnamefont {T.}~\bibnamefont {Padmanabhan}},\ }\bibfield  {title} {\bibinfo {title} {{From gravitons to gravity: Myths and reality}},\ }\href {https://doi.org/10.1142/S0218271808012085} {\bibfield  {journal} {\bibinfo  {journal} {Int. J. Mod. Phys. D}\ }\textbf {\bibinfo {volume} {17}},\ \bibinfo {pages} {367} (\bibinfo {year} {2008})},\ \Eprint {https://arxiv.org/abs/gr-qc/0409089} {arXiv:gr-qc/0409089} \BibitemShut {NoStop}%
\bibitem [{\citenamefont {Deser}(1970)}]{Deser:1969wk}%
  \BibitemOpen
  \bibfield  {author} {\bibinfo {author} {\bibfnamefont {S.}~\bibnamefont {Deser}},\ }\bibfield  {title} {\bibinfo {title} {{Selfinteraction and gauge invariance}},\ }\href {https://doi.org/10.1007/BF00759198} {\bibfield  {journal} {\bibinfo  {journal} {Gen. Rel. Grav.}\ }\textbf {\bibinfo {volume} {1}},\ \bibinfo {pages} {9} (\bibinfo {year} {1970})},\ \Eprint {https://arxiv.org/abs/gr-qc/0411023} {arXiv:gr-qc/0411023} \BibitemShut {NoStop}%
\bibitem [{\citenamefont {Butcher}\ \emph {et~al.}(2009)\citenamefont {Butcher}, \citenamefont {Hobson},\ and\ \citenamefont {Lasenby}}]{Butcher:2009ta}%
  \BibitemOpen
  \bibfield  {author} {\bibinfo {author} {\bibfnamefont {L.~M.}\ \bibnamefont {Butcher}}, \bibinfo {author} {\bibfnamefont {M.}~\bibnamefont {Hobson}},\ and\ \bibinfo {author} {\bibfnamefont {A.}~\bibnamefont {Lasenby}},\ }\bibfield  {title} {\bibinfo {title} {{Bootstrapping gravity: A Consistent approach to energy-momentum self-coupling}},\ }\href {https://doi.org/10.1103/PhysRevD.80.084014} {\bibfield  {journal} {\bibinfo  {journal} {Phys. Rev. D}\ }\textbf {\bibinfo {volume} {80}},\ \bibinfo {pages} {084014} (\bibinfo {year} {2009})},\ \Eprint {https://arxiv.org/abs/0906.0926} {arXiv:0906.0926 [gr-qc]} \BibitemShut {NoStop}%
\bibitem [{\citenamefont {Weinberg}(1964)}]{Weinberg:1964ew}%
  \BibitemOpen
  \bibfield  {author} {\bibinfo {author} {\bibfnamefont {S.}~\bibnamefont {Weinberg}},\ }\bibfield  {title} {\bibinfo {title} {{Photons and Gravitons in $S$-Matrix Theory: Derivation of Charge Conservation and Equality of Gravitational and Inertial Mass}},\ }\href {https://doi.org/10.1103/PhysRev.135.B1049} {\bibfield  {journal} {\bibinfo  {journal} {Phys. Rev.}\ }\textbf {\bibinfo {volume} {135}},\ \bibinfo {pages} {B1049} (\bibinfo {year} {1964})}\BibitemShut {NoStop}%
\bibitem [{\citenamefont {Weinberg}(1965)}]{Weinberg:1965rz}%
  \BibitemOpen
  \bibfield  {author} {\bibinfo {author} {\bibfnamefont {S.}~\bibnamefont {Weinberg}},\ }\bibfield  {title} {\bibinfo {title} {{Photons and gravitons in perturbation theory: Derivation of Maxwell's and Einstein's equations}},\ }\href {https://doi.org/10.1103/PhysRev.138.B988} {\bibfield  {journal} {\bibinfo  {journal} {Phys. Rev.}\ }\textbf {\bibinfo {volume} {138}},\ \bibinfo {pages} {B988} (\bibinfo {year} {1965})}\BibitemShut {NoStop}%
\bibitem [{\citenamefont {Barnich}\ and\ \citenamefont {Henneaux}(1993)}]{Barnich:1993vg}%
  \BibitemOpen
  \bibfield  {author} {\bibinfo {author} {\bibfnamefont {G.}~\bibnamefont {Barnich}}\ and\ \bibinfo {author} {\bibfnamefont {M.}~\bibnamefont {Henneaux}},\ }\bibfield  {title} {\bibinfo {title} {{Consistent couplings between fields with a gauge freedom and deformations of the master equation}},\ }\href {https://doi.org/10.1016/0370-2693(93)90544-R} {\bibfield  {journal} {\bibinfo  {journal} {Phys. Lett. B}\ }\textbf {\bibinfo {volume} {311}},\ \bibinfo {pages} {123} (\bibinfo {year} {1993})},\ \Eprint {https://arxiv.org/abs/hep-th/9304057} {arXiv:hep-th/9304057} \BibitemShut {NoStop}%
\bibitem [{\citenamefont {Deser}\ and\ \citenamefont {Yang}(1990)}]{Deser:1990bk}%
  \BibitemOpen
  \bibfield  {author} {\bibinfo {author} {\bibfnamefont {S.}~\bibnamefont {Deser}}\ and\ \bibinfo {author} {\bibfnamefont {Z.}~\bibnamefont {Yang}},\ }\bibfield  {title} {\bibinfo {title} {{Inconsistency of Spin 4 - Spin-2 Gauge Field Couplings}},\ }\href {https://doi.org/10.1088/0264-9381/7/8/024} {\bibfield  {journal} {\bibinfo  {journal} {Class. Quant. Grav.}\ }\textbf {\bibinfo {volume} {7}},\ \bibinfo {pages} {1491} (\bibinfo {year} {1990})}\BibitemShut {NoStop}%
\end{thebibliography}%

\end{document}